Hydrodynamics of Domain Walls in Multiferroics: Impact on Memory Devices

*Prof. James F. Scott[*]*

Prof. J. F. Scott, School of Chemistry and School of Physics and Astronomy, St. Andrews Univ., St. Andrews, Scotland KY16 9ST
jfs4@st-andrews.ac.uk
Dr. D. M. Evans, Earth Sciences Dept., Cambridge Univ., Cambridge, U. K., CB2 3EQ
Prof. J. M. Gregg, School of Maths. and Physics, Queen's Univ., Belfast, N. I., BT7 1 NN
Prof. A. Gruverman, Dept. Physics and Astron., Univ. Nebraska, Lincoln, NE 68588 USA

1. |Introduction

Very recently and very surprisingly the dynamics of electron transport in both grapheme[1] and some low-temperature metals[2] have been shown to be dominated under some conditions by hydrodynamics. That is, electronic conduction is similar to fluid dynamics, with vortex motion, and not limited by Bloch theory. At about the same time it was shown[3,4] that the ferroelastic domain walls in multiferroics[5] are also controlled by fluid mechanics, with both wrinkling[6] and folding[3] at certain velocity thresholds, and hence that the domain walls may be treated as ballistic objects in high-viscosity media [n.b., wrinkling involves smoothly curved periodic modulation of domain walls, whereas folding consists of nearly 180-degree changes in direction]. The wrinkling-folding instability critical field $E_f$ is known to scale[3b] as the film thickness d as

$$E_f(d) = Ad^{4/9} \qquad (1)$$

but this has not been tested for ferroelastic walls.

## 2. Theoretical Model

2.1 Richtmyer-Meshkov Instabilities

We note that the wrinkling of the $Pb_5Ge_3O_{11}$ domain wall in high applied electric fields E resembles the Richtmyer-Meshkov instability in adjacent fluid bilayers;[7,8] here the field E is suddenly applied and initially produces small amplitude perturbations (wrinkling), which rapidly grow with time to a nonlinear regime (at a threshold of E = 150 kV/cm in lead germanate), resulting in bubble-like injection of spherical nano-ferroelectric +P domains into the –P region (**Figure 1**). The ferroelectric region with polarization P parallel to E behaves as the lighter fluid in the Richtmyer-Meshkov model; if the vertical arrangement of the liquid bilayer is reversed, Meshkov found that needle-structures result, not spherical blobs. This asymmetry seems paradoxical; in fluids it arises from gravity, but in ferroelectrics it is not obvious. The spherical shapes of the nano-domains in Figure 1 are not in themselves evidence of Richtmyer-Meshkov instabilities, since nano-domains are often spherical due to surface tension; the important point is that they are not needle-shaped.

## 3. Experimental

3.1 Forward bias

The data for a positive field (+V in Fig. 1a) are compared ith similar magnetic data, discussed further below, in Fig. 1b. The similarity is striking.

3.2 Reverse bias

In the Richtmyer-Meshkov model, reversing the direction of the applied force (voltage in our case) produces needle-like "domains" rather than spherical blobs. In **Figure 1cd** we show the results of applying a pulse train of alternating +V and –V voltages, with E in each case ca. 200 kV/cm. This results in a superposition of needle-like domains decorated with spherical nano-domains, supporting the prediction of the Richtmyer-Meshkov model. The direction of the needles is not random but favors specific crystallographic axes. In this sense the data differ from those in liquids.

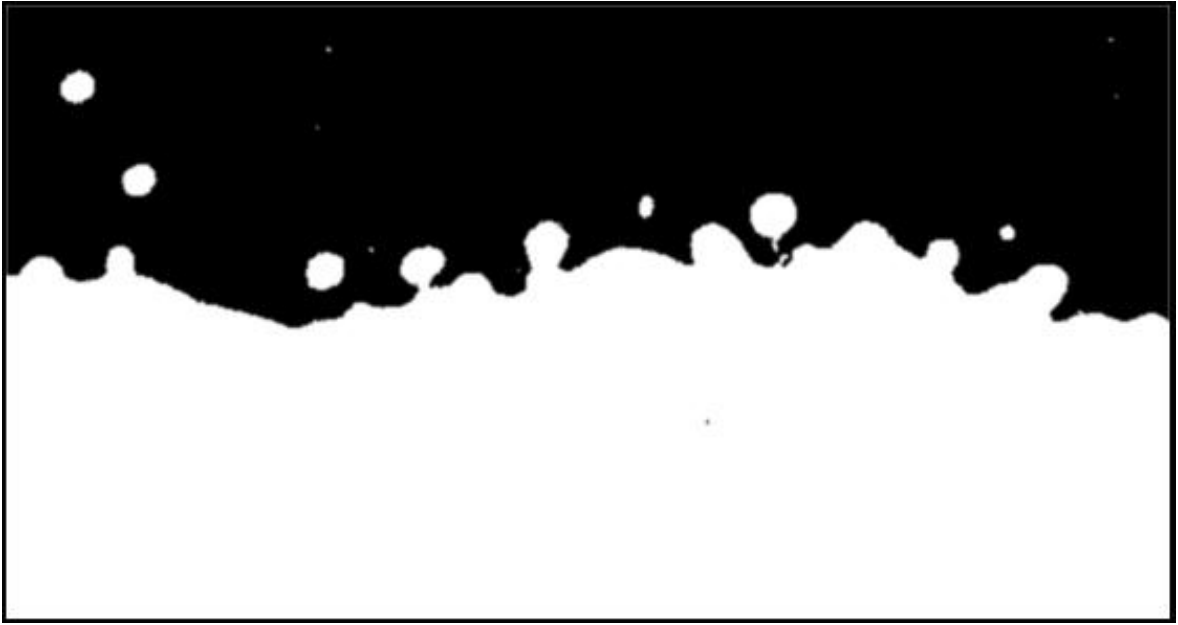

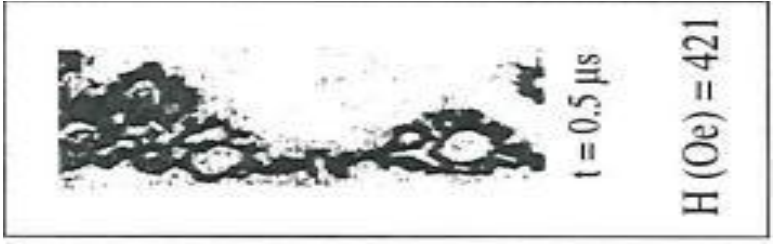
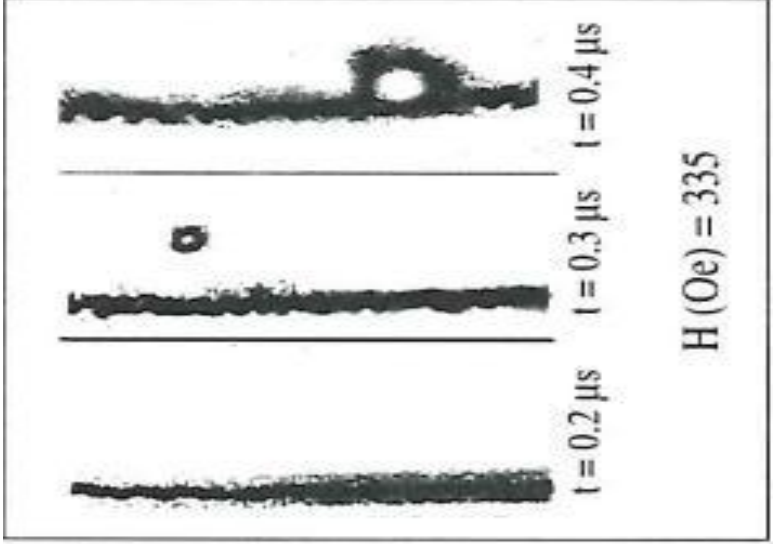
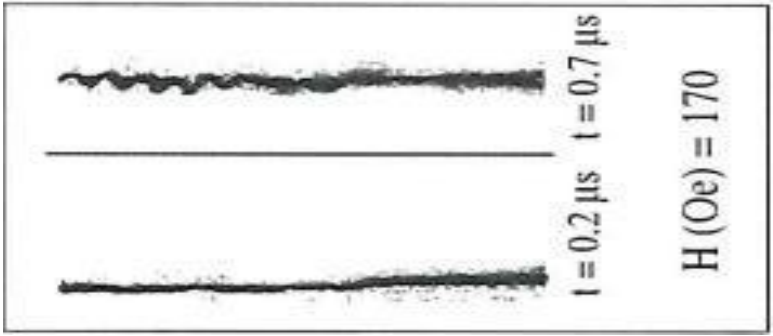

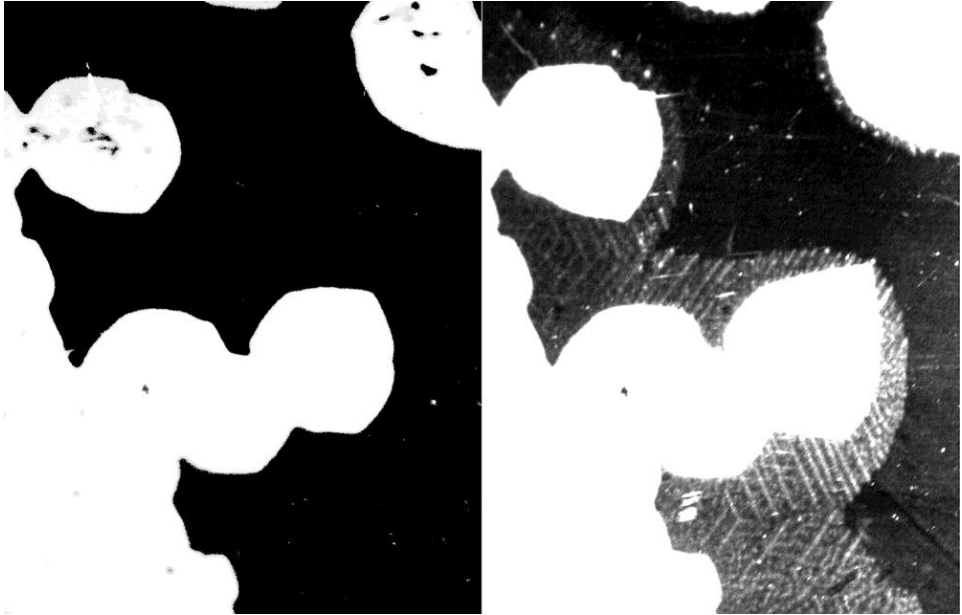

Figure.1 (a). Emission of spherical ferroelectric nanodomains in lead germanate at ca. 150 kVcm-1;[6] (b) Emission of skyrmion-like magnetic nano-domains from a moving magnetic wall at different times and fields; modified from Randoshkin,[19c]; see reference 20a for a matching skyrmion calculation. Not the wrinkling precursors at small fields and short times; (c) Skyrmion-like ferroelectric nano-domains[6,20,21] being ejected upwards from an already wrinkled domain wall [see Reference 3] by an applied electric field of 150 kV/cm at ambient temperature in lead germanate $Pb_5Ge_3O_{11}$ [n.b., this is a large field, compared with the bulk coercive field $E_c$]. Note the longer wavelength wrinkling of the interface. Full width of figure is ca. 35 microns. The lower (white) section is polarized upward; the upper (black) section is polarized downwards. There is a similarity with the patterns of spherical ejections observed in Richtmyer-Meshkov instabilities[7,8] in fluids subjected to nonlinear fields when the lower fluid is less dense than the upper one, whereas if the lower fluid is the denser, needle-like patterns are predicted. (d) Domain pattern for the same specimen in (a) but subjected to a square-wave voltage pulse sequence of +200 kV/cm and -200 kV/cm (left side – no voltage; right side, ac voltage train).

There is also an analogy to voltage-induced crumpling in dielectric membranes, reported very recently.[9] Crumpling is not an exact term but generally refers to stress-driven vertical wrinkles in a horizontal plate (often circular) with a threshold. See reference 10. The use of effective viscosity models is in general not new; it has been used for *magnetic* domains for twenty years.[11a] In addition, the buckling of ferroelastic plates in a magnetic field has also be analyzed,[11b] along with a longer history of nonlinear bifurcations in ferroelastic martensitics[11c] and ferroelectric films.[11d]

In the present work we extend this analysis to relate to quantitative measurements of domain wall velocities[12-15] and effective masses for domain walls.[16]

**3.2 Other Instabilities**

3.2.1 Helfrich-Hurault Mechanism

In contrast to the discussion above of wrinkling and Richtmyer-Meshkov instabilities in ferroelectric-ferroelastic lead germanate, which occur only in high fields E, and where the ferroic walls are simultaneously ferroelectric and ferroelastic, another instability, a purely ferroelastic one, occurs in lead iron tantalate zirconate-titanate at zero electric field but nonzero applied stress; in this system the ferroelectric walls are not coincident with the ferroelastic ones, but lie inside them.

Figure 2 illustrates some unpublished data from reference17. The smallest domains are rectangular ferroelectric domains ca. 5 nm wide which are inside larger (ca. 100-nm) ferroelastic domains. The latter have curved walls.

Our hypothesis is that the ferroelastic walls result from domain motion in a viscous medium (damping is provided by acoustic phonons).[6] Although this provides a plausible explanation for the curved walls, as shown in **Figure 2** below, we emphasize that we have no temporal information to support this dynamical model. Hence it is possible that the curvature shown in Figure1 and 2 arises from some purely electrostatic mechanism.

However, in support of the hydrodynamic viscosity model, we have compatible data and modeling from Salje.[4]

3.2.2 Salje's Model[23]

The basic assumption in Salje's model is that unlike ferroelectric switching, the hysteresis in ferroelastic switching is dominated by continuum fluid mechanics and not the lattice symmetry. He points out that for the strain reversal step from –S to +S, under a positive (reversing) stress to an initially negative strain, the ferroelastic hysteresis is dominated by viscous flow, with a complex domain structure sometimes describable as a "domain glass." By comparison the step from +S(0) to +S(stress) is not hydrodynamic, but proceeds via conventional dynamics and often involves needle-like propagating domains.

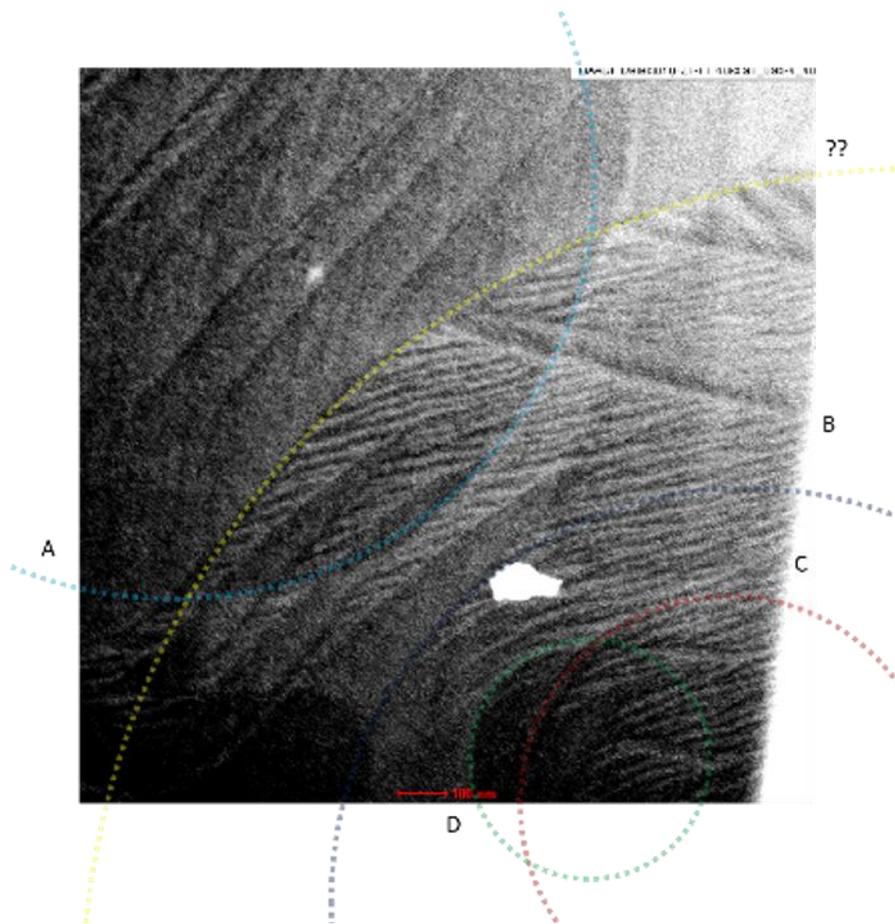

Figure 2

Figure 2. TEM micrograph of lead-iron-tantalate-zirconate-titanate.[17] The ferroelectric domains are black and white stripes ca. 5 nm wide with straight edges, inside at ca. 45 degrees with respect to the walls of the larger ferroelastic domains with curved sides.

White spot is Ga-ion implantation from the FIB process. Parabolic or circular curves fitted as aids to the eye for three walls. Note that the radii of curvature vary considerably, probably ruling out effects of underlying carbon grids.

3.2.4 Open-channel Viscous Flow

Another model we consider here is open channel viscous flow. This is motivated by the curved front such mechanisms produce **(Figure 3)**. It is qualitatively different from the Helfrich-Hurault model above in an important way: It has no velocity threshold. We do not present it here as an alternative to Helfrich-Hurault instability, but rather as a possible low-velocity precursor to the latter, at velocities below the folding threshold.

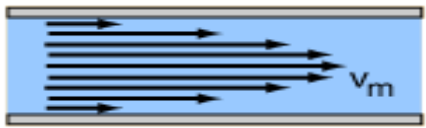

Figure 3. Schematic model of velocity distribution in open-channel viscous flow. This is to be compared with the curved ferroelastic walls in Figure 2. Note that velocity goes to zero at the edges, giving straight edges, as in Figure 2 far from the apparent vertex of each wall; this is a deviation from exactly parabolic.

3.2.5 Preliminary Numerical Parameters

We need to relate the figures above in terms of channel width w, film thickness or depth d, velocity maximum v(m), and some sort of Reynolds number. We can use the creep velocity v = 1 nms$^{-1}$ from Tybell, Paruch and Triscone;[11] see also more recently Ng, Ahluwalia et al.[12,13] and Scott and Kumar.[14] The effective mass for the wall is ca. one proton mass $m_p$.[15]

The basic idea is that relatively low-viscosity media produce folds, whereas higher viscosity media produce skyrmions and vertex structures.[16-19] The vortex structures (and skyrmions[19,20]) are analogous to high viscosity aa-aa lava, just as the smooth folds are analogous to those in lower viscosity ropy pahoehoe. Note also that folding is known to be created via focused ion beams (FIB) in polymers;[21] and the samples in reference 16, including Figure 3 above, were all subject to FIB.

We know a few useful parameters from other work: The average domain wall velocity at low fields in perovskite oxides is ca. 1 nms$^{-1}$.[12-14] The ferroelastic domain wall viscosity is very large compared with normal liquids, and comparable to that in martensitic metals; a rough estimate by Scott[3] is 10$^6$ poise and by Salje and Carpenter[22] is 10$^{13}$ poise, for two different materials. The typical ratio of radius of curvature to domain in-plane width in the narrow direction is ca. 15:1 to 20:1 for the large domains. (The larger the radius of curvature, the larger the shear modulus required for the instability threshold.[31b]) These parameters are may be helpful for future modeling but are insufficient to determine effective Raleigh or Reynolds numbers.

3.2.6 Parabolic Shapes for Ferroelastic Domain Walls

One way to compare data with Helfrich-Hurault boundaries and the shape of open-channel steady-state flow in Figure 3 is the exact shape: Both Helfrich-Hurault boundaries and steady-state fronts are expected to be parabolic, although steady-state fronts should have infinite slopes exactly at the boundaries or pinning sites, where non-slip boundary conditions give zero velocity (Figure 3). Closer inspection of the PFTZT data in Figure 2 shows the longest walls, which have arc lengths of ca. 1.0 microns, are parabolas but have straight edges far away from their apparent vertices.

3.3 Possible artifacts

It is always useful to play Devil's Advocate with one's own data. So we ask whether the TEM data above could arise from the underlying carbon grid in the system. Such grids consist of a connected set of disk-shaped carbon, with radii of curvature similar to those of the larger arcs marked in Figure 2. However, the data shown in Figure 2 (marked) have various different radii of curvature, from $\gg 1$ μm to a little greater than 200 nm, which is not compatible with the carbon grids used. Moreover, our samples are probably too thick (80-100 mm) for carbon grids to be seen through them via TEM. But we do not see identical patterns via atomic force microscopy (AFM/PFM) on samples with no TEM carbon grids. The twisted domains that resemble ropy pahoehoe lava do not match this planar parabolic geometry. Hence, we cannot completely exclude the idea that such carbon grids might form a template (underlying strain pattern) for some of the larger domains illustrated in the figure. Therefore as discussed in Reference 3 there are two possible effects of sample preparation and characterization: The folding of domains may arise from the FIB process; and at least some of the parabolic domain wall shapes may be influenced by the carbon grid in the TEM microscopy. However, we emphasize that the radii of curvature for the ferroelastic domains vary considerably, which is incompatible with a carbon grid TEM pattern.

4. **Summary and Models (Richtmyer-Meshkov and Helfrich-Hurault)**

The observation of nonlinear instabilities for wrinkling and emission of spherical nano-domains in ferroelectrics has previously been described successfully by skyrmion models.[20,21] However, in the present work we broaden that description to make contact with other nonlinear models from fluid mechanics. Evidence for folding in ferroelastic thin films is presented and related to studies on other materials. Such phenomena seem to be rather generic in physics, ranging from gold films[24] to lava.[25] The use of hydrodynamic models in ferroic crystals for elastic domain wall dynamics is however rather new, and the key idea involved for multiferroic PFTZT is that ferroelectric nano-domains lie within ferroelastic micro-domains.[26] This description is qualitatively different from the equilibrium minimum-energy ferroelastic wall model of Roitburd,[27] which is based upon the magnetic domain model of Landau-Lifshitz-Kittel (frequently termed the Kittel Law); the latter models assume infinite lateral dimensions and no folding. They are models based upon minimizing energies at or near equilibrium, not local nonlinear forces. As discussed recently, our preferred models are not close to equilibrium nor small perturbations: Even the old (1943) Ramsberg-Osgood model is of form $F = -kx + bx^2 + c x^n$, where n = ca. 5 and $cx^3 >> b$, not $<< b$. This is not a weak anharmonic perturbation.

There is an extensive literature in the field of liquid crystals on electro-hydrodynamic models, beginning with Gleeson.[28] As in the present case of lead germanate, Gleeson's model for nematics emphasizes a threshold electric field for the onset of convection; however, it was unable to explain the general dendritic fingering patterns observed in nematics, although it showed that folding must be a first-order phase transition. And in addition to the Richtmyer-Meshknov model mentioned above for lead germanate, there is a close analogy for PFTZT with the Helfrich-Hurault layer-instability in liquid crystals.[29-31a] This is a sliding lamellar instability in smectic liquid crystals, characterized, as with the present data, by crescent-shaped domains.[31b] The basic idea there is that hydrodynamic behavior can be described as a power series expansion of the free energies in terms of powers of the gradient tensor components, including terms in $E^2$ and also vorticity (curl x v)$^2$, where v is local velocity. These yield convection thresholds. Unfortunately,[31] no comprehensive theory for such nonlinear irreversible processes exists, nor can the fluctuation-dissipation theorem be used. However, following De Gennes, the critical field $E_c$ for folding can be estimated as $\varepsilon_0 \varepsilon_a E(crit)^2 = 2\pi K/(Ld)$, where L is a length scale equal to the square root of the ratio of shear modulus K to bulk modulus B, and d is a wrinkling length scale of order 100 microns in smectics. This typically gives for smectics with roughly K = 2 pN and B = 50 MNm$^{-2}$, $E_c$ = ca. 200 kV/cm = 20 MV/m, which is close to the wrinkling threshold in Fig.1 for ferroelectric walls in lead germanate of 150 kVcm$^{-1}$. It is important to note that the shapes of domains with these lamellar instabilities is parabolic,[32] similar to that in Figure 2. The similarity of the present layer-instabilities in ferroelastic films and Helfrich-Hurault models should be investigated further. Nonlinear folding in viscous sheets has been studied without simple solutions for more than a century, so one should not underestimate the problem. Recently this task has been taken up by many authors, and theory and experiment for fold sizes are given elsewhere.[33-35] Generally the fold size L varies as the inverse square root of viscosity. For filamentary bifurcations (e.g., honey poured on bread):

$$L = h(4\rho g/3\mu v r^2)^{1/4} \qquad (2)$$

where $\rho$ is density; g, gravitational acceleration; $\mu$, viscosity (10 Pa); r, radius of filament 1 mm); v, velocity (0.1 ms$^{-1}$); and h, height of fall (0.1 m). For honey this gives macroscopic fold lengths ca. a few mm. But for our ferroelastic domain walls, with v = 1 nms$^{-1}$ ($10^8$ smaller[12-15]), viscosity $10^6$-$10^{12}$ times larger,[3,4] and small h, this predicts micron or submicron fold lengths.

Finally we emphasize as far as the curved sides of the ferroelastic domains are concerned, that present data do not discriminate in the case of PFTZT between nonlinear models with velocity thresholds and models such as congealed open-channel flow.

## 5. Implications for Memory Devices

Ferroelectrc memories are already in high-volume production ($100 million/year) level as commercial devices for transit fare cards and cash machines. The chips are produced by Samsung and packaged in Korea and Japan under several brand names, e.g., Felica for fare cards and Edy for cash machines ("e-money"). The active ferroelectric material is also ferroelastic. If these memories are to widen their applications to faster devices, then higher fields will be required (same voltage but thinner films). At present such devices typically run at 5V across 100 nm (50 kVcm$^{-1}$); hence the instability here near 150 kVcm$^{-1}$ is only a factor of x3 above present norms. At such fields the ferroelastic wall instability thresholds discussed here may serve as the rate-limiting parameters.

Acknowledgements:

Received:
Revised:
Published online: